\documentclass[twocolumn,prl,aps]{revtex4}
\usepackage[dvipdfm]{graphicx}
\usepackage{dcolumn}
\usepackage{bm}
\usepackage{color}
\usepackage{ulem}

\begin{document}
\newcommand{\bR}{\mbox{\boldmath $R$}}
\newcommand{\tr}[1]{\textcolor{red}{#1}}
\newcommand{\trs}[1]{\textcolor{red}{\sout{#1}}}
\newcommand{\tb}[1]{\textcolor{blue}{#1}}
\newcommand{\tbs}[1]{\textcolor{blue}{\sout{#1}}}
\newcommand{\Ha}{\mathcal{H}}
\newcommand{\mh}{\mathsf{h}}
\newcommand{\mA}{\mathsf{A}}
\newcommand{\mB}{\mathsf{B}}
\newcommand{\mC}{\mathsf{C}}
\newcommand{\mS}{\mathsf{S}}
\newcommand{\mU}{\mathsf{U}}
\newcommand{\mX}{\mathsf{X}}
\newcommand{\sP}{\mathcal{P}}
\newcommand{\sL}{\mathcal{L}}
\newcommand{\sO}{\mathcal{O}}
\newcommand{\la}{\langle}
\newcommand{\ra}{\rangle}
\newcommand{\ga}{\alpha}
\newcommand{\gb}{\beta}
\newcommand{\gc}{\gamma}
\newcommand{\gs}{\sigma}
\newcommand{\vk}{{\bm{k}}}
\newcommand{\vq}{{\bm{q}}}
\newcommand{\vR}{{\bm{R}}}
\newcommand{\vQ}{{\bm{Q}}}
\newcommand{\vga}{{\bm{\alpha}}}
\newcommand{\vgc}{{\bm{\gamma}}}
\newcommand{\Ns}{N_{\text{s}}}
\newcommand{\avrg}[1]{\left\langle #1 \right\rangle}
\newcommand{\eqsa}[1]{\begin{eqnarray} #1 \end{eqnarray}}
\newcommand{\eqwd}[1]{\begin{widetext}\begin{eqnarray} #1 \end{eqnarray}\end{widetext}}
\newcommand{\hatd}[2]{\hat{ #1 }^{\dagger}_{ #2 }}
\newcommand{\hatn}[2]{\hat{ #1 }^{\ }_{ #2 }}
\newcommand{\wdtd}[2]{\widetilde{ #1 }^{\dagger}_{ #2 }}
\newcommand{\wdtn}[2]{\widetilde{ #1 }^{\ }_{ #2 }}
\newcommand{\cond}[1]{\overline{ #1 }_{0}}
\newcommand{\conp}[2]{\overline{ #1 }_{0#2}}
\newcommand{\nn}{\nonumber\\}
\newcommand{\cdt}{$\cdot$}
\newcommand{\bra}[1]{\langle#1|}
\newcommand{\ket}[1]{|#1\rangle}
\newcommand{\braket}[2]{\langle #1 | #2 \rangle}
\newcommand{\bvec}[1]{\mbox{\boldmath$#1$}}
\newcommand{\blue}[1]{{#1}}
\newcommand{\bl}[1]{{#1}}
\newcommand{\bn}[1]{\textcolor{blue}{#1}}
\newcommand{\rr}[1]{{#1}}
\newcommand{\bu}[1]{\textcolor{blue}{#1}}
\newcommand{\red}[1]{{#1}}
\newcommand{\fj}[1]{{#1}}
\newcommand{\green}[1]{{#1}}
\newcommand{\gr}[1]{\textcolor{green}{#1}}
\definecolor{green}{rgb}{0,0.5,0.1}
\definecolor{blue}{rgb}{0,0,0.8}
\preprint{APS/123-QED}

\title{
Mott Physics on
Helical Edges of 2D Topological Insulators
}
\author{Youhei Yamaji}
\email{yamaji@solis.t.u-tokyo.ac.jp}
\author{Masatoshi Imada}
\affiliation{Department of Applied Physics, University of Tokyo, Hongo, Bunkyo-ku, Tokyo, 113-8656, Japan.}%
\affiliation{JST-CREST, Hongo, Bunkyo-ku, Tokyo, 113-8656, Japan.
}%
\date{\today}

\begin{abstract}
{We study roles of electron correlations on topological insulators
on the honeycomb lattice with the spin-orbit interaction. Accurate variational Monte Carlo calculations 
show that   
the increasing on-site Coulomb interactions 
cause a strong suppression of the charge Drude weight in the helical-edge metallic states leading to a presumable Mott transition from a conventional topological insulator to an edge Mott insulator before a transition to a bulk antiferromagnetic insulator.
The intermediate bulk-topological and edge-Mott-insulator phase has a helical spin-liquid character with the protected time-reversal symmetry.    
}
\end{abstract}
\pacs{
}
\maketitle
\paragraph{-{\it Introduction.}}
Recently,
spin Hall insulators and its generalization, topological insulators
(TIs) have attracted much attention as a new state of matter\cite{Kane_Mele}.
A remarkable feature of the newly discovered quantum phase
is the $Z_2$-type topological distinction from other conventional phases as well as
the existence of robust gapless edges or surface states concomitant with the bulk 
insulating gap, which
are all protected by the time reversal (TR) symmetry.
The edge or surface modes of TI provide us with
truly one- or two-dimensional gapless and metallic electronic states.

It has also been proposed that
TI may appear in systems 
under substantial electron correlations 
such as in 4$d$ or 5$d$ transition metal oxides\cite{Young08,Pesin_Balents10,Yang10,Wan10,Shitade09}, while the interplay of electron 
correlations with the topological insulator has not been well understood,
although the absence of the back scattering protected by the time reversal symmetry
is expected to suppress 
electron correlation effects\cite{Kane_Mele,Wu_Bernewig_Zhang06,Xu_Moore06}. 

In this Letter, based on results of calculations obtained from 
a multi-variable variational Monte Carlo (MVMC) methods improved by 
Tahara and one of the authors\cite{Tahara08}, 
we propose that electron correlation effects introduced by an onsite
interaction, namely, a Hubbard $U$ 
in the Kane-Mele model on the honeycomb lattice
allow a transition from the above TI to an unconventional TI phase characterized by the charge 
gapful (insulating) but spin gapless edge excitations with a nonzero spin Drude weight
within the same preserved topological nontriviality of
the bulk states that are protected by the time reversal symmetry. 
This new topological edge Mott insulator (TEMI) phase  
is stabilized in a region of the intermediate correlation strength
sandwiched by a bulk antiferromagnetic insulator (BAFI)  with the broken time reversal symmetry (or bulk Mott insulator (BMI)) 
in the larger $U$ region and the simple TI insulator in the weak correlation region.

\begin{figure}[h]
\begin{center}
\includegraphics[width=7cm]{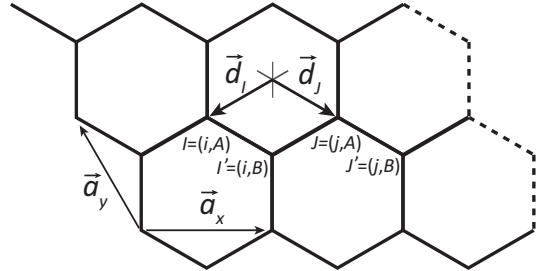}
\end{center}
\caption{
Honeycomb lattice on which HKM model is defined.
\label{Fig_HC}}
\end{figure}

\paragraph{-{\it Model.}}
We study a tight binding hamiltonian on
the two-dimensional honeycomb lattice proposed
by Kane and Mele~\cite{Kane_Mele}
with inclusion of the on-site Coulomb interaction, and without the Rashba term
to study electron correlation effects on the topological
insulator.
Hereafter we call this simple model the Hubbard-Kane-Mele model and is defined as
\eqsa{
	\hat{\mathcal{H}}
	=
	\hat{\mathcal{H}}_{{\rm KM}}
	+U\sum_{I}\hat{n}_{I\uparrow}\hat{n}_{I\downarrow},
}
with 
\eqsa{
   \hat{\mathcal{H}}_{{\rm KM}}
	=-t\sum_{\langle I,J \rangle\sigma}\hatd{c}{I\sigma}\hatn{c}{J\sigma}
	+it_{2}\sum_{\langle\langle I,J \rangle\rangle \alpha \beta}
	\nu_{ij}\hatd{c}{I\alpha}[\sigma_{z}]_{\alpha\beta}\hatn{c}{J\beta}
}
where $\hat{\mathcal{H}}_{{\rm KM}}$ is the Kane-Mele hamiltonian and
$U$ generates an onsite Hubbard interaction between the up and down spin electrons.
Here we define
$\nu_{ij}=\vec{d}_{i}\times\vec{d}_{j}/\left| \vec{d}_{i}\times\vec{d}_{j}\right|$,
and $I=(i,a)$ $(a = A,B)$
(see Fig.\ref{Fig_HC}).

\paragraph{-{\it Method.}}
We perform unrestricted Hartree-Fock (UHF) calculations as well as MVMC calculations by optimizing a large number of variational
parameters.

In the UHF calculation, we decouple the $U$ term as 
\eqsa{
&\hat{n}_{I\uparrow}\hat{n}_{I\downarrow}\simeq
\avrg{\hat{n}_{I\uparrow}}\hat{n}_{I\downarrow}
+
\avrg{\hat{n}_{I\downarrow}}\hat{n}_{I\uparrow}
-
\avrg{\hat{n}_{I\uparrow}}\avrg{\hat{n}_{I\downarrow}}
\nonumber\\
&-
\langle \hatd{c}{I\uparrow}\hatn{c}{I\downarrow}\rangle\hatd{c}{I\downarrow}\hatn{c}{I\uparrow}
-
\langle \hatd{c}{I\downarrow}\hatn{c}{I\uparrow}\rangle\hatd{c}{I\uparrow}\hatn{c}{I\downarrow}
+
\langle\hatd{c}{I\uparrow}\hatn{c}{I\downarrow}\rangle\langle\hatd{c}{I\downarrow}\hatn{c}{I\uparrow}\rangle
\label{Eq:UHF}
}
For the MVMC calculations, 
we employ a variational wave function~\cite{Tahara08} defined as
\begin{equation}
|\psi\ra = \sP_{\rm G}\sP_{\rm J}|\phi_{\rm pair}\ra,
\label{Eq:WF}
\end{equation}
where $\sP_{\rm G}$ is the Gutzwiller factor defined by
\begin{equation}
\sP_{\rm G}=\exp [-\sum_{I}g_{I}\hat{n}_{I\uparrow}\hat{n}_{I\downarrow}],
\label{Eq:GF}
\end{equation}
and $\sP_{\rm J}$ is the Jastrow factor
defined by
\begin{equation}
\sP_{\rm J}=\exp \left[-\frac{1}{2}\sum_{I,J}v_{IJ}(\hat{n}_{I\uparrow}+\hat{n}_{I\downarrow})
(\hat{n}_{J\uparrow}+\hat{n}_{J\downarrow})\right],
\label{Eq:JF}
\end{equation}
{with the spatially dependent variational parameters $g_I$ and $v_{IJ}$.}
We impose the Gutzwiller factor on all the sites, whereas introduce the Jastrow factor only along
the zig-zag edges.
The one-body part $|\phi_{\rm pair}\ra$ is
a
 generalized pairing wave function defined as
\begin{equation}
|\phi_{\rm pair}\ra=\Big[\sum_{i,j=1}^{\Ns}f_{ij}c_{i\uparrow}^{\dag}c_{j\downarrow}^{\dag}\Big]^{N/2} |0 \ra
\end{equation}
with $f_{ij}$ being the complex variational parameters.
In this study, we 
allow $f_{ij}$ to have
2-sublattice ($2\times L_{y}$-sublattice) structure
or equivalently we have $2\times 2\times N_{\rm s}$
($2\times L_{y}\times 2 \times L_{y} \times L_{x}$) variational parameters for the torus (cylinder).
All the variational parameters are simultaneously 
optimized by using the stochastic reconfiguration method~\cite{Tahara08,Sorella01}
generalized for complex variables.
The accuracy of this method has been proven in a number of 
benchmarks\cite{Tahara08,Imada_Miyake10}

Charge and spin Drude weights are calculated by introducing vector potentials as
the Peierls factors,
\eqsa{
	t_{IJ\sigma}\rightarrow t_{IJ\sigma}\exp [i \vec{A}_{\sigma}\cdot \vec{r}_{IJ}],
}
where
$\vec{r}_{I}=n_{I}\vec{a}_{x}+m_{I}\vec{a}_{y}$ and $\vec{r}_{IJ}=\vec{r}_{I}-\vec{r}_{J}$.
Here $n_{I}$ and $m_{J}$ are integers, and lattice vectors are $\vec{a}_{x}$ and $\vec{a}_{y}$
(see Fig.\ref{Fig_HC}).
From this Peierls factor, the charge and spin Drude weights are calculated from the energy stiffness
\begin{equation}
D_{{\rm c}}=
\left.
\frac{1}{2}\frac{d^2 E (\vec{A}_{\uparrow},\vec{A}_{\downarrow})}{d |\vec{A}|^2}
\right|_{\vec{A}_{\uparrow}=\vec{A}_{\downarrow}}
\label{Drude}
\end{equation} 
and
\begin{equation}
D_{{\rm s}}=
\left.
\frac{1}{2}\frac{d^2 E (\vec{A}_{\uparrow},\vec{A}_{\downarrow})}{d |\vec{A}|^2}
\right|_{\vec{A}_{\uparrow}=-\vec{A}_{\downarrow}}
\end{equation} 
To clarify the
edge state, we employ a cylinder with sizes $N_{s}$=$L_x$$\times$$L_y$$\times$2, for the honeycome lattice with two sites on a unit cell 
and the periodic (free) boundary conditions
in the $x$ ($y$) directions.
We have confirmed that the 
employed 
width $L_y$ is large enough to make isolated two edges at the two free boundaries at $y=0$ and $y=L_y$.
For the bulk properties we employ the torus, where the boundary is periodic for all the directions .  

\begin{figure}[h]
\begin{center}
\includegraphics[width=8cm]{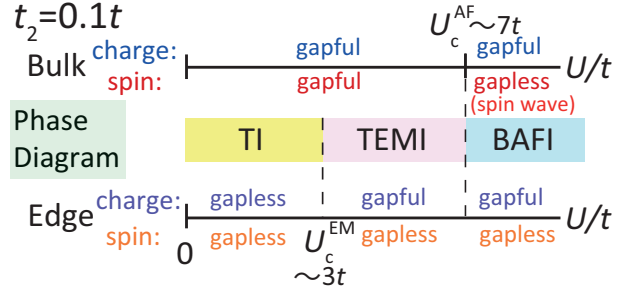}
\end{center}
\caption{
(color online).
Phase diagram of HKM {model} 
obtained by MVMC containing three phases for $t_{2}=0.1t$.
The bulk on the torus shows
a phase transition from TI to the bulk antiferromagnetic insulator (BAFI) 
at $U/t\sim 7$.
On the cylinder,
the edge shows insulating behaviors
for 
$3t\lesssim U$.
The intermediate TEMI is a gapless spin liquid at the edges.
\label{Fig_phase}}
\end{figure}
\begin{figure}[h]
\begin{center}
\includegraphics[width=7.5cm]{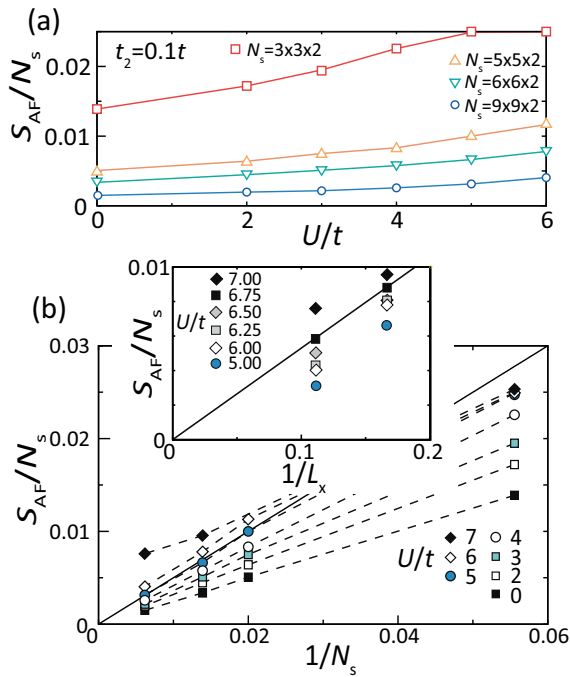}
\end{center}
\caption{
(color online).
(a) $U$-dependence of $S_{{\rm AF}}(\vec{q}=\vec{0})$
for various system {sizes}.
(b) $1/N_s$ dependences of 
$S_{{\rm AF}}(\vec{q}=\vec{0})$.
The inset shows {the same quantity vs. $1/\sqrt{N_s/2}\ (=1/L_x$)}.
We note that $S_{AF}/N_s$ may be scaled linearly with $1/N_s$ in the asymptotic region ($1/N_s\rightarrow 0$) of the disordered phase, as we see in the main panel of Fig. 3(b), whereas as is known in the spin wave theory, it should be scaled linearly with $1/L_x$ with a nonzero offset in the asymptotic region of the ordered phase, as we see in the inset of Fig.3(b).
These scalings suggest that
the phase transition occurs between $U=6.75t$ and $U=7t$.
Error bars are all within the symbol sizes.
}
\label{Fig_Sq}
\end{figure}

\paragraph{-{\it Bulk phase diagram.}}
The ground state phase diagram of the bulk 
is shown in Fig.\ref{Fig_phase} for the HKM model on the torus.
Our MVMC results show the antiferromagnetic transition at $U=U_{{\rm c}}^{{\rm AF}}\sim 7t$
for $t_{2}=0.1t$.
Below $U\sim 7t$, the bulk stays as a topological insulator and the peak height of the magnetic structure factor 
{defined by 
\eqsa{
	S_{{\rm AF}}(\vec{q})
	=\frac{1}{3N_{s}}\sum_{I,J}e^{i\vec{q}\cdot\vec{r}_{ij}}
	\lambda_{I}\lambda_{J}
	\vec{\hat{S}}_{I}\cdot\vec{\hat{S}}_{J},
}
for the spin-1/2 operator $\vec{\hat{S}}_{I}$ scales to a size-independent constant after the size extrapolation in contrast to the Bragg peak height proportional to 
{$N_{s}=L_{x}\times L_{y}\times 2$} observed in BAFI,
as is shown in Fig.\ref{Fig_Sq}.
Here we have shown the peak values, which appear at the wavenumber $q=0$ and for the
{staggered} modes within the unit cell,
namely $\lambda_{I}=+1$(-1) for $I=(i,A)$ ($I=(i,B)$).
}
The MVMC calculation gives the critical value of $U$ for $t_{2}\neq 0$
larger than that for $t_{2}=0$.
These results are consistent with
an auxiliary-field quantum Monte Carlo simulation for $t_2$=0, which
shows $U_{{\rm c}}^{{\rm AF}}$=4.3$t$\cite{Meng}.

Here the magnetic moments in the BAFI phase align in the $xy$-plane.
This fact conforms with the UHF results (see also Ref.\onlinecite{Rachel}) and
the effective hamiltonian at
the strong coupling limit;
the second order perturbation of the second-neighbor hopping $it_{2}$ yields
the second-neighbor exchange coupling as
$J_{2}[\hat{S}_{i}^z\hat{S}_{j}^z-\hat{S}_{i}^x\hat{S}_{j}^x
-\hat{S}_{i}^y\hat{S}_{j}^y]$, where $J_2 = 4t_{2}^2/U$.
In the $xy$-plane, $J_{2}$ gives
the ferromagnetic coupling and 
stabilizes the BAFI moment within this plane.
The non-zero magnetic moment within the $xy$-plane
always opens a gap at the edge modes,
although an infinitesimal magnetic moment along the $z$-axis
does not open a gap.
\paragraph{-{\it Edge transport.}}
We show MVMC results for the Drude weights
for the HKM model on the cylindrical geometry with two zig-zag edges
along the $x$-direction.
A nonzero Drude weight $D_c$ defined in (\ref{Drude}) represents the coherent charge transport or metalicity\cite{Kohn}.
If we introduce spin-dependent vector potentials,
$\vec{A}_{\sigma}=\sigma\vec{A}$,
we obtain the Drude weight for the spin channel\cite{Kopietz}, namely the spin Drude
weight.
A 
direct evaluation of
the 
$Z_2$ topological number proposed by Lee and Ryu\cite{Lee}
could be done by using the same procedure.
However, it requires much more computational cost, and is left for future studies.

\begin{figure}[h]
\begin{center}
\includegraphics[width=9cm]{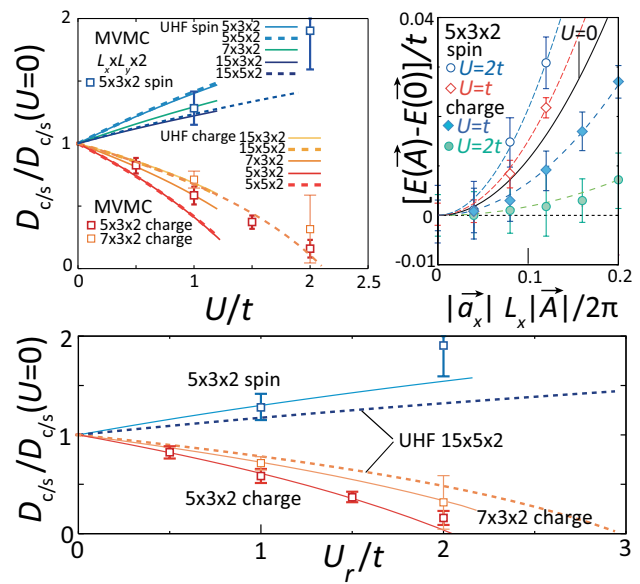}
\end{center}
\caption{
(color online).
{Upper left panel: $U$ dependence of renormalized Drude weights. Upper right panel: Vector potential
dependences of total energy $E$
for $L_x =5$, $L_y =3$.
Lower panel:
Renormalized $D_c(U_r/t)/D_c(0)$ of the UHF and MVMC. The solid curves for the UHF data scales identically with the MVMC data for the same sizes when we employ the renormalized interaction $U_r$.
}
}
\label{Fig_Drude}
\end{figure}

In Fig.\ref{Fig_Drude},
we compare the results for charge and spin Drude weights by MVMC with those of the UHF approximation. 
The data for $15 \times 5 \times 2$ well represent the thermodynamic limit of the nonmagnetic UHF solution.
We see consistent suppression (enhancement) of the charge (spin) Drude weight $D_c (D_s)$ arising from the increasing on-site Coulomb interaction $U$.
Moreover, the suppression (enhancement) of $D_c (D_s)$ has 
nearly
linear dependences on $U/t$.  
{
The renormalized dependences of $D_c(U_r/t)/D_c(0)$ vs. $U_r/t$ with $U_r=U$ for MVMC and $U_r\simeq 1.4\times U$ for UHF make all of them universal,
indicating that the thermodynamic limit for the MVMC is extracted from the UHF at the same $U_r$
as is shown in the lower panel of Fig.\ref{Fig_Drude}.
The MVMC results support a transition on the edge from the TI to a charge gapful (insulating) phase (TEMI) around $U_{c}^{{\rm EM}}\sim 3t$
in the thermodynamic limit. 
}

The suppression and enhancement in the Drude weights are naturally accounted by focusing on the spin and charge pumping caused by the vector potentials. The spin-independent (spin-dependent) vector potential causes the spin (charge) pumping along the zig-zag edges\cite{Kane_Mele}, which is nothing but the celebrated quantum spin Hall effects in the Kane-Mele model. Without the Rashba term, which mixes the spin-up and -down components, the spin-independent vector potential
{causes} a spin pumping of the $z$-components. Here we note that the small amount of spin accumulation of the $z$-component does not induce a gap opening at the edge modes.

The spin pumping generates spin polarization along the $z$-axis
at the edges, which helps electrons to reduce the cost of the Hubbard $U$.
{Contrarily},
the charge pumping forces to increase the double occupation
at the edges resulting in the cost of $U$.
Therefore, the energy increase with increasing vector potential (namely the stiffness or the Drude weight as the quadratic coefficients) 
decreases (increases) compared with $E$ of the non-interacting system.
Such spin-charge separated Drude weights appear even in the restricted Hartree-Fock
calculation that does not allow the magnetic moments within the $xy$-plane.
\paragraph{-{\it Edge phase diagram.}}
Based on the calculated  spin and charge Drude weights, 
we show the edge phase diagram in Fig.\ref{Fig_phase}. It supports a
metal-insulator transition at the edge at 
{$U=U_{{\rm c}}^{{\rm EM}} \sim 3t$,
}
where the bulk still continues to be paramagnetic TI beyond it.
In contrast, 
the coherent edge spin transport is enhanced by increasing $U$.

The low-energy effective model, namely,
the helical Tomonaga-Luttinger liquid (HTLL)
does not include the back scattering and Umklapp channels\cite{Wu_Bernewig_Zhang06,Xu_Moore06},
which are, in general, essential for the formation of insulating phases.
However,
for $t_2 =0.1t$ and 
{$U\sim 3t$}, the amplitude of the bulk gap,
which limits the energy scale to justify the treatment by the topological
band insulator,
becomes comparable with $U$.
Then, HTLL
will fail to capture this
{Mott} insulating behaviors of the edge modes.
In fact,
a large Coulomb repulsion ($U$ $\gg$ $t$, $t_{2}$)
inevitably prohibits coherent propagations of
electrons.
In reality, it actually
results in gapful charge excitations
for $t_{2}=0.1t$ and 
{$U_{{\rm c}}^{{\rm EM}}(\sim 3t) \leq U \leq U_{{\rm c}}^{{\rm AF}}(\sim 7t)$
}
 with gapless spin excitations
at the edge for the HKM.

\paragraph{-{\it Whole phase diagram}}
{As we see in Fig.\ref{Fig_phase}, through 
the $U$ variation, 
the bulk state is always insulating (charge gapful) if the spin-orbit interaction is nonzero
while it has a spin gapless excitation exclusively in the conventional BMI or BAFI phase at the largest $U$ region.}
However, the edge state is always characterized by the gapless spin excitations while the charge excitation is gapless only in the lowest $U$ region of the TI phase.
Then we find the intermediate phase, TEMI where the bulk excitations are gapful in both spin and charge channels, while in the edge state the charge excitation is gapful (insulating) and the spin excitation is gapless with a spin liquid behavior. 
We note this general phase diagram with three phases contained may be universal also in three-dimensional systems except for the additional possibility that, depending on the lattice geometry, the spin liquid in the TEMI phase could be replaced with the magnetic symmetry breaking such as the antiferromagnetic order at the edge (surface) as is proposed by ref.\cite{Shitade09}.

\paragraph{-{\it Discussion}}
A possible interpretation of the TEMI phase 
is a spinon liquid under the fractionalized electrons in the slave-rotor approximation proposed by Young {\it et al.}\cite{Young08}, and Pesin and Balents\cite{Pesin_Balents10}.  However, the 1D spin-charge separation in the simple Hubbard chain is characterized by the spin/charge excitations as bosonic collective modes of the Tomonaga-Luttinger liquid\cite{Schulz,Voit}.  In the present helical spin liquid on the edges, it is also likely to have gapful charge and gapless spin  bosonic collective density modes leading to the {HTLL}\cite{Wu_Bernewig_Zhang06,Xu_Moore06} distinct from the chiral Tomonaga-Luttinger liquid \cite{Voit} in the quantum Hall phase. 
   
The conventional TI in the cylinder geometry is characterized by the nonzero diagonal charge (spin) conductivities denoted by $\sigma_{ccxx}\ne 0$ ($\sigma_{ssxx}\ne 0$)  and nonzero spin-charge transverse conductivity denoted by $\sigma_{csxy}=\sigma_{scxy}\ne 0$ and $\sigma_{csyx}=\sigma_{scyx}\ne 0$, where all are solely from the edge contributions. Other spin-charge off-diagonal elements are zero.  On the other hand, the present TEMI with the gapful charge and gapless spin liquid edges 
{keeps $\sigma_{ssxx}\ne 0$ again contributed only from the edge, 
whereas all the other elements including $\sigma_{ccxx}$, $\sigma_{csxy}=\sigma_{scxy}$ and $\sigma_{csyx}=\sigma_{scyx}$ vanish. 
The Onsager reciprocal relation of course always holds.
In the both TI and TEMI phases, all the bulk conductivities vanish, while      
in the BMI (or BAFI) phase, 
}
the bulk and edge spin conductivities may remain nonzero with all the other linear responses involving the charge part vanish irrespective of bulk or edge.

\paragraph{-{\it Summary}}
Our present variational Monte Carlo calculations show that the local electron correlation $U$ of the Hubbard-Kane-Mele model on the honeycomb lattice drives a strong crossover or a quantum phase transition within the topologically nontrivial phase. The transition appears to separate an edge metallic TI phase at lower $U$ from TEMI phase with charge gapful and spin gapless (spin liquid) edges at larger $U$, where a bulk charge-spin gap is always retained through these two topological phases. Namely, the larger $U$ phase is characterized by a vanishing {charge} Drude weight together with a nonzero and large spin Drude weight in contrast to the both large charge and spin Drude weights in the lower $U$ phase.  With further {increase of} $U$, this TEMI phase undergoes a transition into the BAFI phase with {the} time reversal symmetry breaking.       
\paragraph{-{\it Acknowledgements}}
Y.Y thanks D. Tahara for sharing his VMC code for real $f_{ij}$.
He also thanks T. Misawa, H. Shinaoka, and M. Kurita
for useful discussions.
We thank financial support from Computational Materials Science Initiative,
and MEXT Japan under the grant number 22104010.

\end{document}